\documentclass{article}
\usepackage{amsmath, amssymb, amsfonts}
\title{ON A STATIONARY SPINNING STRING SPACETIME}
\author{Hristu Culetu \\Ovidius University, Dept.of Physics \\B-dul Mamaia 124, 8700 Constanta, Romania \\e-mail : hculetu@yahoo.com}

\begin{document}
\numberwithin{equation}{section}
\pagenumbering{arabic}
\maketitle
\newcommand{\fv}{\boldsymbol{f}}
\newcommand{\tv}{\boldsymbol{t}}
\newcommand{\gv}{\boldsymbol{g}}
\newcommand{\OV}{\boldsymbol{O}}
\newcommand{\wv}{\boldsymbol{w}}
\newcommand{\WV}{\boldsymbol{W}}
\newcommand{\NV}{\boldsymbol{N}}
\newcommand{\hv}{\boldsymbol{h}}
\newcommand{\yv}{\boldsymbol{y}}
\newcommand{\RE}{\textrm{Re}}
\newcommand{\IM}{\textrm{Im}}
\newcommand{\rot}{\textrm{rot}}
\newcommand{\dv}{\boldsymbol{d}}
\newcommand{\grad}{\textrm{grad}}
\newcommand{\Tr}{\textrm{Tr}}
\newcommand{\ua}{\uparrow}
\newcommand{\da}{\downarrow}
\newcommand{\ct}{\textrm{const}}
\newcommand{\xv}{\boldsymbol{x}}
\newcommand{\mv}{\boldsymbol{m}}
\newcommand{\rv}{\boldsymbol{r}}
\newcommand{\kv}{\boldsymbol{k}}
\newcommand{\VE}{\boldsymbol{V}}
\newcommand{\sv}{\boldsymbol{s}}
\newcommand{\RV}{\boldsymbol{R}}
\newcommand{\pv}{\boldsymbol{p}}
\newcommand{\PV}{\boldsymbol{P}}
\newcommand{\EV}{\boldsymbol{E}}
\newcommand{\DV}{\boldsymbol{D}}
\newcommand{\BV}{\boldsymbol{B}}
\newcommand{\HV}{\boldsymbol{H}}
\newcommand{\MV}{\boldsymbol{M}}
\newcommand{\be}{\begin{equation}}
\newcommand{\ee}{\end{equation}}
\newcommand{\ba}{\begin{eqnarray}}
\newcommand{\ea}{\end{eqnarray}}
\newcommand{\bq}{\begin{eqnarray*}}
\newcommand{\eq}{\end{eqnarray*}}
\newcommand{\pa}{\partial}
\newcommand{\f}{\frac}
\newcommand{\FV}{\boldsymbol{F}}
\newcommand{\ve}{\boldsymbol{v}}
\newcommand{\AV}{\boldsymbol{A}}
\newcommand{\jv}{\boldsymbol{j}}
\newcommand{\LV}{\boldsymbol{L}}
\newcommand{\SV}{\boldsymbol{S}}
\newcommand{\av}{\boldsymbol{a}}
\newcommand{\qv}{\boldsymbol{q}}
\newcommand{\QV}{\boldsymbol{Q}}
\newcommand{\ev}{\boldsymbol{e}}
\newcommand{\uv}{\boldsymbol{u}}
\newcommand{\KV}{\boldsymbol{K}}
\newcommand{\ro}{\boldsymbol{\rho}}
\newcommand{\si}{\boldsymbol{\sigma}}
\newcommand{\thv}{\boldsymbol{\theta}}
\newcommand{\bv}{\boldsymbol{b}}
\newcommand{\JV}{\boldsymbol{J}}
\newcommand{\nv}{\boldsymbol{n}}
\newcommand{\lv}{\boldsymbol{l}}
\newcommand{\om}{\boldsymbol{\omega}}
\newcommand{\Om}{\boldsymbol{\Omega}}
\newcommand{\Piv}{\boldsymbol{\Pi}}
\newcommand{\UV}{\boldsymbol{U}}
\newcommand{\iv}{\boldsymbol{i}}
\newcommand{\nuv}{\boldsymbol{\nu}}
\newcommand{\muv}{\boldsymbol{\mu}}
\newcommand{\lm}{\boldsymbol{\lambda}}
\newcommand{\Lm}{\boldsymbol{\Lambda}}
\newcommand{\opsi}{\overline{\psi}}
\renewcommand{\tan}{\textrm{tg}}
\renewcommand{\cot}{\textrm{ctg}}
\renewcommand{\sinh}{\textrm{sh}}
\renewcommand{\cosh}{\textrm{ch}}
\renewcommand{\tanh}{\textrm{th}}
\renewcommand{\coth}{\textrm{cth}}

\begin{abstract}

~The properties of a stationary massless string endowed with intrinsic spin are discussed. The spacetime is Minkowskian geometrically but the topology is nontrivial due to the horizon located on the surface $r = 0$, similar with Rindler's spacetime.

~For r less than the Planck length b, $g_{\varphi \varphi}$ has the same sign as $g_{tt}$ and closed timelike curves are possible. We assume an elementary particles' spin originates in the frame dragging phenomenon produced by the rotation of the source. 

The Sagnac time delay is calculated and proves to be constant. 
\end{abstract}



\section{INTRODUCTION}
~~It is a known fact that the physical interpretation of the parameters from the vacuum solutions of Einstein's equations with cylindrical symmetry is still a matter of discussion \cite {WB} \cite {HPS}. A reason might be the fact that some parameters are related to topological defects which have no contribution to the Riemann tensor (Einstein's equations give informations about the geometry, not the topology of spacetime).

The spacetime outside an axially symmetric source, endowed with angular momentum was found by Lewis \cite {TL}, solving the Einstein equations in vacuum. Besides the usual dragging effect due to a rotating source, a topological frame dragging appears \cite{HS}, related to the topological defect associated with some metric parameters, which are responsable for the angular momentum of a spinning string (a thin tube of false vacuum).

Mazur \cite{PM} has raised the problem of the quantization of the energy of the particles in the background of a spinning cosmic string (with vanishing mass per unit length), which is considered as a gravitational analog of the Aharonov - Bohm solenoid. 

Ruggiero \cite{MR} and Rizzi and Ruggiero \cite {RR} introduced a gravito - magnetic Aharonov - Bohm effect and formally derived the Sagnac effect by analogy, both in flat and curved spacetimes. The Sagnac time delay was computed for matter or light beams counter-propagating on a round trip, in a rotating frame in flat spacetime, for Kerr, Godel and Schwarzschild geometries. Ruggiero \cite {MR} found that the phase shift is nonvanishing even when the observer is not rotating, due to the angular momentum of the source. 

Deser and Jackiw \cite {DJ} and Deser, Jackiw and 't Hooft \cite {DJH} consider the case of a stationary string carrying mass and an intrinsic spin J per unit length. They focused on the fact that an infinitely long, thin cosmic string supports closed timelike curves (CTC) when $J\neq 0$, but the mass is related to the deficit angle. Their line element is locally Minkowskian because the presence of the string is hidden. It could be uncovered by an "improper" coordinate transformation which mixes the time- and angular- coordinates (the new time variable jumps by $8 \pi G J/ c^{4}$ whenever the string is circumvented).

 We treat in this paper only the case of a massless spinning string. Although the geometry is flat locally, the angular momentum of the source leads globally to a nontrivial topology (the analogy with the case of a uniformly accelerated Rindler observer is apparent \cite {CD}). We show that the spacetime has a horizon on the surface $r = 0$ due to the nondiagonal form of the metric. In addition, the modulus of the (imaginary) horizon surface gravity is 1/b, where b is considered to be of the order of the Planck length. We also conjecture that the elementary particles' spin originates in the frame dragging effect due to the rotation of the source. 
 
 At $r = b$ from the string, $g_{\varphi \varphi}$ changes sign, the $\varphi$ - coordinate becomes temporal and CTC may appear. It is interesting to notice that the Sagnac time delay is constant, irrespective of the enclosed area by the counter-propagating light or matter beams. Moreover, it is nonzero even though the angular velocity of the source of light vanishes.
 
 From now on we choose the geometrical units $G = c = \hbar = 1$.

\section{THE SPINNING STRING METRIC}

 Let us consider the general line element for a cylindrical symmetric stationary spacetime
\begin{equation}
ds^{2} = -f d t^{2} + 2k dt d \varphi + e^{\mu} (d r^{2} + d z^{2}) + l d \varphi^{2}
\label{2.1}
\end{equation}
where f, k, $\mu$, l are functions of the radial coordinate r (we take $x{^0}=t, x^{1} =r, x^{2} =z, x^{3} = \varphi )$. 

 Solving Einstein's field equations in vacuum, we have \cite {HS}
\begin{equation}
f = \alpha r^{-n+1} - \frac{\beta^{2}}{\alpha n^{2}} r^{n+1} ~~,~~~ k = -a f
\label{2.2}
\end{equation}
\begin{equation}
e^{\mu} = r^{(n^{2}-1)/2}, ~~~~ l = \frac{r^{2}}{f} - a^{2} f,~~~a = \frac{\beta r^{n+1}}{n \alpha f} + b.
\label{2.3}
\end{equation}
~The real constants n, $\alpha, \beta$, b are related to the physical characteristics of the model (see \cite {SHPS} \cite {UC}). The parameter n is associated with the Newtonian mass per unit length of a uniformly line mass while $\alpha$ is connected to an arbitrary constant gravitational potential from the corresponding Newtonian solution. The parameters $\beta$ and b are responsable for the nonstaticity of the metric (with $\beta = b = 0$, the geometry (2.1) becomes the static Levi - Civita metric \cite {WS}. b is directly related to the angular momentum of a spinning string and produces a topological frame dragging.
$\beta$ measures the vorticity of the source.

It is worth to note that the Riemann tensor is nonzero only for $n \neq 1$. For $n = 1$ the spacetime (2.1) is flat locally.

~We are dealing in this paper only with a particular case : $n = \alpha =1~~ and~~ \beta =0$. More general cases will be treated elsewhere. Therefore, eqs. (2.2) and (2.3) yield $f = 1, a = b, k = -b~~ and~~ l = r^{2} - b^{2}$. The spacetime (2.1) appears now as 
\begin{equation}
ds^{2} = -d\tau^{2} = -dt^{2} + dr^{2} + dz^{2} - 2 b dt d\varphi + (r^{2} - b^{2}) d\varphi^{2}
\label{2.4}
\end{equation}
The geometry (2.4) may be obtained from the Minkowski metric, written in cylindrical coordinates
\begin{equation}
ds^{2} = -d\bar{t}^{2} + d\bar{r}^{2} + d \bar{z}^{2} + \bar{r}^{2} d\bar{\varphi}^{2}
\label{2.5}
\end{equation}
by means of the coordinate transformation \cite {DJ}
\begin{equation}
t = \bar{t} - b \varphi,~~~r = \bar{r},~~~z = \bar{z},~~~\varphi = \bar{\varphi}
\label{2.6}
\end{equation}
From now on we shall take b of the order of the Planck length $l_{P} = 10^{-33} cm$. That corresponds to $J = \hbar/4 l_{P}$ in the Deser - Jackiw paper.

We could say that the transformation (2.6) generates a topological defect which makes (2.4) not to be equivalent globally with the flat matric (2.5). Even though both spacetimes are flat, the variable t jumps by $2 \pi b$ after a complete rotation around the string. Moreover, for $r < b$ (very close to the axis of rotation), CTCs are possible.

\section{HORIZON AND SURFACE GRAVITY}

We look now for a Killing horizon (stationary null surface) in the geometry (2.4), invariant under time translation. Keeping in mind that the metric is not diagonal, the horizon is obtained from the condition \cite {CGLP} 
\begin{equation}
\hat{g}_{00} \equiv g_{00} - \frac{g_{03}^{2}}{g_{33}} = 0
\label{3.1}
\end{equation}
which leads to $r_{H} = 0$. The angular velocity of the horizon, with respect to a nonrotating observer at infinity, is
\begin{equation}
\Omega_{H} = -\frac{g_{03}}{g_{33}}|_{r = 0} = -\frac{1}{b}
\label{3.2}
\end{equation}
The region $g_{33} = r^{2} - b^{2} < 0$ or $r < b$ leads to CTC (time machine region) and the boundary $r = b$ is the velocity of light surface. Because $g_{03} \neq 0$ when $g_{03} = 0$, the metric is nonsingular. In addition, $r = b$ is a timelike surface : the timelike curves may cross into the time machine and viceversa.

 The expression for the surface gravity $\kappa$ of the horizon \cite {CGLP} is
\begin{equation}
\kappa^{2} = \nabla_{\mu}L \nabla^{\mu}L
\label{3.3}
\end{equation}
where 
\begin{equation}
L^{2} = -g_{00} -2\Omega_{H}g_{03} - \Omega_{H}^{2}g_{33}
\label{3.4}
\end{equation}
One obtains $L^{2} = -r^{2}/b^{2}$, which leads to $\kappa^{2} = -1/b^{2}$, i.e. $\kappa$ is imaginary, with $|\kappa| = 1/b$. The reason may be related to the fact that the horizon is located inside the time machine region. In spite of its imaginary value, an analogy with the surface gravity $g$ of the Rindler horizon at $\ro = 0$
\begin{equation}
ds^{2} = - g^{2} \rho^{2} dt^{2} + d\rho^{2} + dy^{2} + dz^{2} 
\label{3.5}
\end{equation}
is apparent. In both situations we have only one parameter at our disposal : g, respectively b in the case of the spinning string. Therefore, the surface gravities are g and $|\kappa|$, respectively.

\section{THE SAGNAC EFFECT}

Let us consider now two light beams counter-propagating on a circular trajectory (using a system of mirrors) in a rotating reference frame, in flat spacetime \cite {MR} \cite{RR}. Due to the rotation of the source of light, we know that a phase shift and, from here, a time delay is measured (Sagnac effect) between the two counter-propagating beams. Using an analogy with the expressions for the Coriolis and Lorentz forces, Rizzi and Ruggiero \cite {RR} showed that the effect is also valid for matter beams : Cooper pairs, charged particles, etc. 

Ruggiero \cite {MR} proved that even when observer's angular velocity is vanishing, there is a time delay in curved spacetime, when the source of the gravitational field has a nonvanishing angular momentum. Even though the spacetime (2.4) is flat locally, we show that a Sagnac time delay is present because of the spinning string (the Deser - Jackiw ''cosmon'' \cite {DJH}, if we supress the z-coordinate). 
\\With $\dot{r} \equiv dr/d\tau = 0$ and $\dot{z} = 0$, eq. (2.4) gives for the light beam
\begin{equation}
1 + 2~b~\frac{d\varphi}{dt} - (r^{2} - b^{2}) (\frac{d\varphi}{dt})^{2} = 0
\label{4.1}
\end{equation}
whence
\begin{equation}
\frac{d\varphi}{dt} \equiv \Omega_{\pm} = \frac{1}{\pm R - b}
\label{4.2}
\end{equation}
where $R = const.$ is the radius of the circular orbit. $\Omega_{+}~ and ~\Omega_{-}$ corresponds to the two counter-rotating beams. We see that outside the time machine $(R > b)$, $\Omega_{+} > 0 ~and~\Omega_{-} < 0$. In addition, when $b = 0$ is taken (the Minkowski form of the metric), we have $|\Omega_{+}| = |\Omega_{-}|$ and no phase shift is detected. On the horizon $r = 0,~ \Omega_{+} = \Omega_{-} = \Omega_{H} = -1/b$, as expected.

Let us compute now the Sagnac time delay. We have
\begin{equation}
\Omega_{+} = \left(\frac{d\varphi}{dt}\right)_{+} = \frac{1}{R-b},~~~\Omega_{-} = \left(\frac{d\varphi}{dt}\right)_{-} = \frac{-1}{R+b}
\label{4.3}
\end{equation}
~Therefore
\begin{equation}
t_{+} = 2 \pi (R-b),~~~~t_{-} = 2 \pi (R+b)
\label{4.4}
\end{equation}
whence
\begin{equation}
\Delta t = t_{-} - t_{+} = 4 \pi b
\label{4.5}
\end{equation}
In other words, the time delay does not depend upon the area enclosed by the photon's trajectory. \footnote {That might be a consequence of the fact that $g_{03} = -b = const.$. Even though the observer who measures the time delay does not rotate, $\Delta t \neq 0$ due to the angular momentum of the source. We assume that the spinning string (thin tube of false vacuum) is localized on the axis of rotation}

In the first order in $b/R$, $\Omega_{+}~~ and~~ \Omega_{-}$ may be written as
\begin{equation}
\Omega_{+} \approx \frac{1}{R} + \frac{b}{R^{2}} ,~~~~\Omega_{-} \approx -\frac{1}{R} + \frac{b}{R^{2}}
\label{4.6}
\end{equation}
~We see that far from the source $(R >> b)$ the term $b/R^{2}$ in $\Omega_{+}$ and $\Omega_{-}$ plays the same role as the angular velocity $\omega$ of a uniformly rotating observer in Minkowski spacetime \cite {DS}. In other words, the spinning source ''rotates'' the surroundings with an angular velocity $b/R^{2}$ (the frame dragging phenomenon). \footnote {With, for example, $R \approx 10^{-11} cm,~ b/R^{2} \approx 0.3 s^{-1}$ and the first term $1/R \approx 3.10^{21} s^{-1}$. Therefore, $\Omega_{\pm} \approx (\pm 3.10^{21} + 0.3) s^{-1}$}. The effect of the spinning string is the same as if the source of the light beam rotated with the angular velocity $0.3 s^{-1}$ on a circle of radius $10^{-11} cm$. The shorter the radius R, the greater the 2-nd term of (4.6).

It is worthwhile to notice that, in the approximation used, $b/R^{2}$ equals ''the angular velocity of the local nonrotating observer'' $\Omega$ (nonrotating relative to the local spacetime geometry), given by 
\begin{equation}
\Omega \equiv \frac{1}{2} \left(\Omega_{+} + \Omega_{-} \right) = - \frac{g_{03}}{g_{33}} = \frac{b}{R^{2}-b^{2}}
\label{4.7}
\end{equation}

~Keeping in mind that there is no structure along the z - axis \cite {DJ}, we could eliminate that direction. The spinning string becomes now a ''point particle'' (the Deser - Jackiw ''cosmon''), located on the $z = 0$ plane. In that case, the intrinsic spin per unit length J might be interpreted as ''the intrinsic spin per Planck length''. Therefore, the angular momentum L of the ''point cosmon'' is $L = l_{P} J$. But in our model $J \approx b/l_{P}$, whence $L \approx \hbar$. On the grounds of the above argument, we conjecture that the origin of elementary particles' spin is due to the dragging effect of the rotating cosmon
\footnote {It is a known fact that ''a spin $1/2$ particle has an orientation entanglement with its environment'' \cite {MP}. That is valid when the particle is considered to be rigid, whose action contains the square of the extrinsic curvature of the particle's world line}, located at $r = 0$ (the spacetime horizon).

\section{CONCLUSIONS}

~~Several properties of the Deser - Jackiw spinning string, viewed as a thin tube of false vacuum, have been discussed in this paper. The spacetime is flat only locally, the difference compared to Minkowski geometry being topological (a comparison with the Rindler metric was given). Even though $g_{00} = -1$, the spacetime possesses a horizon at $r = 0$ due to the nondiagonal form of the line element.

~Because of the nonvanishing angular momentum of the string, a phase shift will appear between two counter circumnavigating light or matter beams, the time delay being constant and proportional to the Planck length b. The frame dragging effect was proposed to be at the origin of the elementary particles' spin (something similar to an orientation entanglement with its environment, valid for a spin 1/2 particle). We also assumed the spinning string is localized on the axis of rotation. In our view, the rotating system is the agent who localizes the string.\\

ACKNOWLEDGMENTS\\
~I would like to thank the Organizers of the NEB XII Conference, ''New Developments in Gravity'' for inviting me to attend this very stimulating meeting where the paper was presented and for their warm hospitality during my stay at Nafplio.

\end{document}